\documentclass[aps,pra,twocolumn]{revtex4}
\usepackage{graphicx}
\usepackage{bm}
\usepackage{amsmath}
\usepackage[colorlinks=true,allcolors=blue]{hyperref}

\providecommand{\bra}[1]{\langle #1 \rvert}
\providecommand{\ket}[1]{\lvert #1 \rangle}

\providecommand{\be}{\begin{equation}}
\providecommand{\ee}{\end{equation}}
\providecommand{\ba}{\begin{eqnarray}}
\providecommand{\ea}{\end{eqnarray}}

\usepackage{mathbbol}
\usepackage[makeroom]{cancel}

\usepackage{xcolor}

\begin{document}

\title{Quantum versus classical transport of energy in coupled two-level systems}

\author{I. Medina}
\affiliation{Centro de Ci\^encias Naturais e Humanas, Universidade Federal do ABC (UFABC), Santo Andr\'e, SP, 09210-580, Brazil}
\author{S. V. Moreira}
\affiliation{Centro de Ci\^encias Naturais e Humanas, Universidade Federal do ABC (UFABC), Santo Andr\'e, SP, 09210-580, Brazil}
\author{F. L. Semi\~ao}
\affiliation{Centro de Ci\^encias Naturais e Humanas,  Universidade Federal do ABC (UFABC), Santo Andr\'e, SP, 09210-580, Brazil}

\begin{abstract}
We consider the problem of energy transport in a chain of coupled quantum systems with the goal of shedding light on how nonclassical resources can affect transport.
We study the cases for which either coherent or incoherent energy hopping takes place in the chain. Here, incoherent energy hopping is referred to as the ``classical'' scenario in allusion to its fully diagonal dynamics in the basis formed by the eigenstates of the decoupled sites. We focus on the case of a linear chain of two-level sites and find a hopping rate threshold above which the coherent quantum case is more efficient than the incoherent counterpart. We then link the quantum hopping rate to the coherence global maximum, which allows us to state that there is a coherence threshold above which the quantum scenario is more efficient. Next, we consider the integrated coherence generated by the dynamics and show how it is related to what is known as the invasiveness of a quantum operation.  Our results strongly suggest the significant role played by quantum invasiveness as a resource for quantum transport.

\end{abstract}
\vskip2pc

\maketitle

\section{Introduction}

Since the experiments performed by Aspect {\it et al.} in the 1980s, which claimed the violation of a Bell inequality using entangled-polarized photons \cite{aspect1,aspect2},  
quantum entanglement have been in the spotlight when nonclassicality is discussed  \cite{Horodecki}.
As a natural development, in recent years we have witnessed the emergence of several other nonclassicality indicators beyond entanglement, which are also relevant for the full understanding of quantum phenomena.  In the context of the present study, we will focus on those which have recently been systematized under the resource theoretic framework \cite{Coecke}, namely coherence \cite{Baumgratz} and quantum invasiveness \cite{Moreira4}.
Naturally, with the development of quantum technologies, it has become crucial to investigate how nonclassical resources are related to the efficiency of a certain task \cite{Cimini, Streltstov, Adesso, rqchannel,  Kurashvili}. Such connections would help us to learn how nonclassical resources could be employed to fully exploit quantum technologies. 
One important phenomenon for which these studies seem to be relevant is energy transport.


Understanding the phenomenon of energy transport in quantum systems is a very relevant and timely research topic.
Examples of coupled systems where nonclassical phenomena are important to energy transfer are numerous, including highly complex systems strongly coupled to their environment, from molecular aggregates in photosynthetic complexes \cite{PlenioTransport,Guzik} to polymeric samples \cite{poly}.
Since the seminal work reporting the experimental observation of quantum dynamics in the energy transport inside a given photosynthetic complex \cite{wavelike}, the number of studies dedicated to the quantum description of transport has sharply increased \cite{Schachenmayer, Semiao, Feist,EngVibAssEnTranfIons,fmo1,fmo2,fmo3,EnvoiAssTranspIonsMaier, NMTransport,Elinor,Cao}.
Here, we aim to shed some light on the relationship between the phenomenon of energy transfer in coupled two-level quantum systems and resources such as coherence \cite{Baumgratz} and quantum invasiveness \cite{Moreira4}.
To do so, classical scenarios are defined and modeled in a sound way in terms of the {\it classical} operations that may allow energy transfer to occur in an incoherent way between the coupled sites of a chain.
In turn, coherent coupling is established between the sites of the chain in quantum scenarios.
By using a coherence quantifier, we first analyze how our approach allows one to characterize the appearance of quantum advantage in terms of coherence in a specific model.
Furthermore, we will see that the framework of the resource theory of invasiveness of quantum operations \cite{Moreira4} captures the classical and quantum scenarios defined above in a natural way.
By finding a quantifier of quantum invasiveness in connection with coherence, we will show how this nonclassical resource provided by quantum operations can also be related to the efficiency of quantum transport. It is worth mentioning that quantum operations as resources for quantum tasks are a timely and active topic of research \cite{rqchannel,Plenio1,Plenio2,Saxena,Takagi,Li}.

This paper is organized as follows. 
First, we present the transport model and the classical and quantum scenarios in Sec. \ref{RQT}.
Then, in Sec. \ref{discussion}, we present our results. 
By exploring the system dynamics, we study examples where the quantum advantage in the efficiency of energy transport  manifests.
Then, we briefly review the resource theoretic framework of quantum invasiveness to explore its connection with transport efficiency by finding a suitable quantifier of this nonclassical resource of quantum operations. 
We finish this section with a presentation of a possible setup where our ideas can be experimentally assessed. In Sec. \ref{Conclusions}, we present our final remarks.

\section{Transport Model}\label{RQT}
\begin{figure}[h!]
	\includegraphics[scale=0.3]{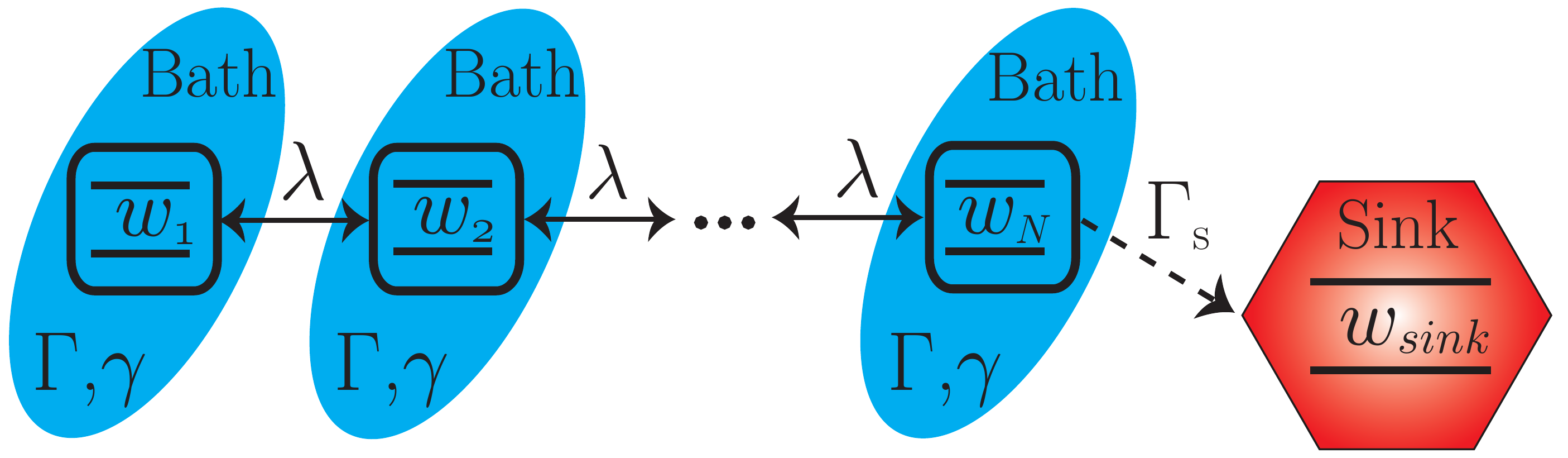}
	\caption{(Color online) Transport scenario considered in this work. It is described by a linear chain of $N$ coupled two-level systems (sites). The coupling strength between the sites is given by the parameter $\lambda$. Each site is also subjected to a local environment which causes  {\it dephasing} and spontaneous emission under rates $\gamma$ and $\Gamma$, respectively. The last site is incoherently coupled to a sink, where the energy is collected.
	\label{scheme} }
\end{figure}

As mentioned before, energy transport is an important feature of coupled molecular systems, such as photosynthetic complexes \cite{lhs1,lhs2,lhs3} and organic photovoltaic cells \cite{solarcell}.
Its essential features are captured by the model depicted in Fig. \ref{scheme}, which consists of a linear chain of $N$ first-neighbor coupled two-level systems (sites).

The system Hamiltonian $H = H_F + H_I$ consists of a free part ($\hbar=1$),
\begin{equation}\label{FreeH}
H_F=\frac{\omega}{2}\sum_j\sigma_j^z,
\end{equation}
and a part which accounts for the coupling between first neighbours in the chain
\begin{align}
H_I=\lambda\sum_{j=1}^{N-1}\left(\sigma_{j}^+\sigma_{j+1}^-+\sigma_{j}^-\sigma_{j+1}^+\right).
\label{Qcoupling}
\end{align}
In these equations, $\sigma^z_j$ is the Pauli $z$ operator in the basis $\{\ket{g}_j,\ket{e}_j\}$ with $\ket{e}_j$ ($\ket{g}_j$) being the excited (ground) state of site $j$, and $\omega$ is the energy associated with each two-level system. Also,  $\sigma_j^-=\ket{e}_j\bra{g}=(\sigma_j^+)^\dag$ are ladder operators and $\lambda$ is a coupling constant considered real without loss of generality.

The initial state of the system is fixed by putting one excitation in first site of the chain and no excitation in all other sites.
To evaluate the transport efficiency, we consider the $N$th site of the chain to be dissipating into an auxiliary two-level system, known as the sink $s$.
Also, we consider that each site of the chain is subjected to local dissipation and local dephasing to account for the presence of noise.
We note that the results presented below turn out to be independent of the local frequencies. 
However, the assumption of local noise is valid only in experimental implementations for which $\lambda\ll\omega$ \cite{Santos, Hofer, Gonzalez, Mitchison, McConnella, DeChiara}. Later, we will provide an example of a concrete setup where this condition is fulfilled.
 In this scenario, the noise in each site $j$ is described by the Lindblad superoperator
\begin{align}
\mathcal{L}_j(\rho)=\Gamma(2\sigma_j^+\rho\sigma_j^- -\sigma_j^-\sigma_j^+\rho-\rho\sigma_j^-\sigma_j^+)+\gamma(\sigma^z_j\rho\sigma^z_j - \rho),
\end{align}
where $\Gamma$ and $\gamma$ are the dissipation and dephasing rates, respectively. 
In turn, the coupling between the $N$th site and the sink is described by 
\begin{align}
\mathcal{L}_{\rm sink}(\rho)=&\Gamma_s(2\sigma_N^-\sigma_s^+\rho\sigma_s^-\sigma_N^+\nonumber\\
&-\sigma_s^-\sigma_N^+\sigma_N^-\sigma_s^+\rho-\rho\sigma_s^-\sigma_N^+\sigma_N^-\sigma_s^+),
\label{incohcoup}
\end{align}
where $\Gamma_s>0$ is the rate of energy transferred to the sink. 
Therefore, the system dynamics is described by
\begin{align}
\frac{\partial \rho}{\partial t}=-i[H,\rho]+\mathcal{L}_{\rm sink}(\rho)+\sum_{j=1}^{N}\mathcal{L}_j(\rho).
\label{qdynamics}
\end{align}
We now define a central concept in our work which is the {\it classical transport scenario}. As no coherence in the basis of $H_F$ is to be created, we replace the coherent coupling  $H_I$ by an incoherent term described by a Lindblad operator with the ``same intensity'' of $H_I$, i.e., the same coupling strength $\lambda$. This is described by
\begin{align}\label{totalL}
\mathcal{L}_{C}(\rho)=\lambda[\mathcal{L}_{R}(\rho)+\mathcal{L}_{L}(\rho)]
\end{align}
where
\begin{align}
\mathcal{L}_{R}(\rho)=\sum_{j=1}^{N-1}&(2\sigma_j^-\sigma_{j+1}^+\rho\sigma_{j+1}^-\sigma_{j}^+\nonumber\\
-&\sigma_{j+1}^-\sigma_j^+\sigma_j^-\sigma_{j+1}^+\rho-\rho\sigma_{j+1}^-\sigma_j^+\sigma_j^-\sigma_{j+1}^+),
\end{align}
and 
\begin{align}
\mathcal{L}_{L}(\rho)=\sum_{j=1}^{N-1}&(2\sigma_j^+\sigma_{j+1}^-\rho\sigma_{j+1}^+\sigma_{j}^-\nonumber\\
-&\sigma_{j+1}^+\sigma_j^-\sigma_j^+\sigma_{j+1}^-\rho-\rho\sigma_{j+1}^+\sigma_j^-\sigma_j^+\sigma_{j+1}^-).
\end{align}
Physically, this mechanism corresponds to thermally activated energy migration between nearest neighbors \cite{ion-channel}. 
Then, the classical transport scenario dynamics is the result of the master equation,
\begin{align}
\frac{\partial \rho_C}{\partial t}=-i[H_F,\rho_C]+\mathcal{L}_{C}(\rho_C)+\mathcal{L}_{\rm sink}(\rho_C)+\sum_{j=1}^{N}\mathcal{L}_j(\rho_C),
\label{cdynamics}
\end{align}
where the subscript in $\rho_C$ reminds us that the system state remains classical for all times, i.e., a convex sum of the eigenstates of $H_F$. From this point on, we shall use the subscripts $Q$ and $C$ to distinguish between quantities computed using the quantum or classical scenarios.

It is important to emphasize that we are not stating that Eq. (\ref{cdynamics}) alone is the classical equation of motion used to describe situations like electron transfers between donor and acceptor chemical species. In such complex situations, classical and quantum methods must take into account features like the vibrational degrees of motion and polarization properties of the solvent, for instance. When calling Eq. (\ref{cdynamics})  the classical dynamics, we only meant that it is not able to create coherence or state superpositions in the basis of $H_F$, in contrast to the role played by $H_I$ in Eq. (\ref{qdynamics}).

To investigate and compare the transport efficiency in both scenarios, we study $P_Q(t)$ and  $P_C(t)$, which are the sink population at time $t$ as evaluated with Eq.(\ref{qdynamics}) and Eq.(\ref{cdynamics}), respectively. Their asymptotic values will be called efficiency \textit{per se} and will be denoted $\eta_Q$ and $\eta_C$ \cite{PlenioTransport},
\begin{eqnarray}
\eta_Q &=&P_Q(\infty)=\lim_{t\rightarrow\infty}{\rm{Tr}}[\rho_Q|e\rangle_s\langle e|],\nonumber\\
\eta_C &=&P_C(\infty)=\lim_{t\rightarrow\infty}{\rm{Tr}}[\rho_C|e\rangle_s\langle e|].
\end{eqnarray}
First, we want to investigate the quantum case in light of the coherence it manifests.  The coherence of a quantum state $\rho$, in a certain basis, will be quantified by the relative entropy of coherence \cite{Baumgratz}, defined as
\begin{align}
C(\rho)=S(\rho_{\text{diag}})-S(\rho),
\label{relatentropy}
\end{align}	
where $S(\rho)={\rm Tr}[-\rho{\rm ln}(\rho)]$ is the von Neumann entropy and $\rho_{\text{diag}}$ denotes the diagonal state obtained from $\rho$ by erasing all the off-diagonal elements.\\

\section{Results}\label{discussion}

To illustrate the framework presented in the previous section, we start with numerical simulations of Eqs. (\ref{qdynamics}) and (\ref{cdynamics}) using $N=3$ plus one sink. Later on, we increase $N$ while keeping the scenario of small controllable chains. For simplicity, we keep $\Gamma_s=1.0$ a.u. (arbitrary units) unless otherwise stated. 

\subsection{System dynamics and efficiency}
 
\begin{figure}[t!]
	\includegraphics[scale=0.3]{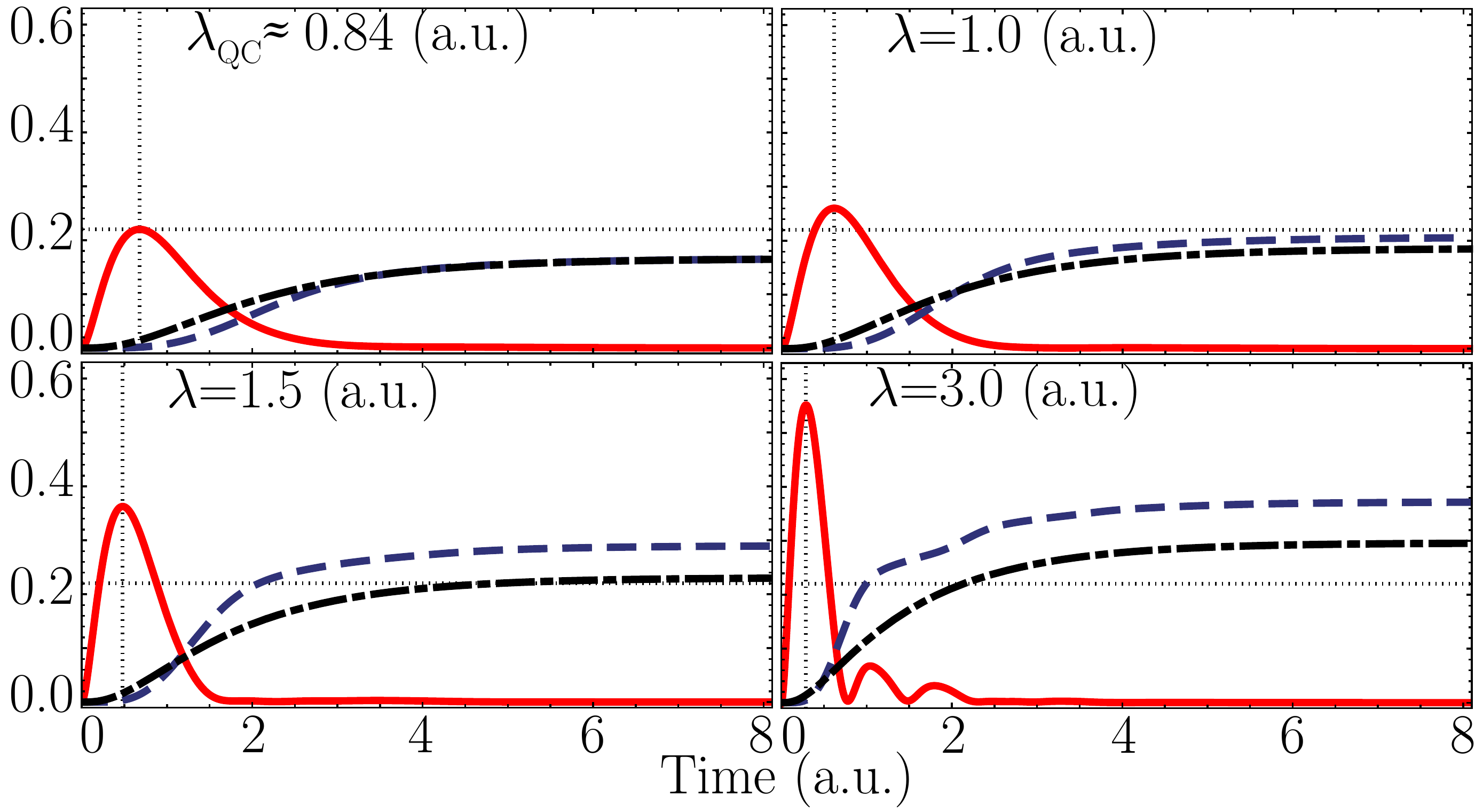}
	\caption{(Color online) \label{dynamics} Plots of $P_C(t)$ (black-dot-dashed line), $P_Q(t)$ (blue-dashed line) and $C(\rho_Q)$ (red-solid line) as a function of time. For all plots we fixed $\Gamma=0.5$ a.u and $\gamma=0.25$ a.u. while the values of $\lambda$ are shown in each panel. The horizontal dotted line represents the maximum coherence when $\lambda=\lambda_{QC}$. The insets show the dynamics of $P_C(t)$ and $P_Q(t)$ for very short times.}
\end{figure}

\begin{figure}[t!]
	\includegraphics[scale=0.41]{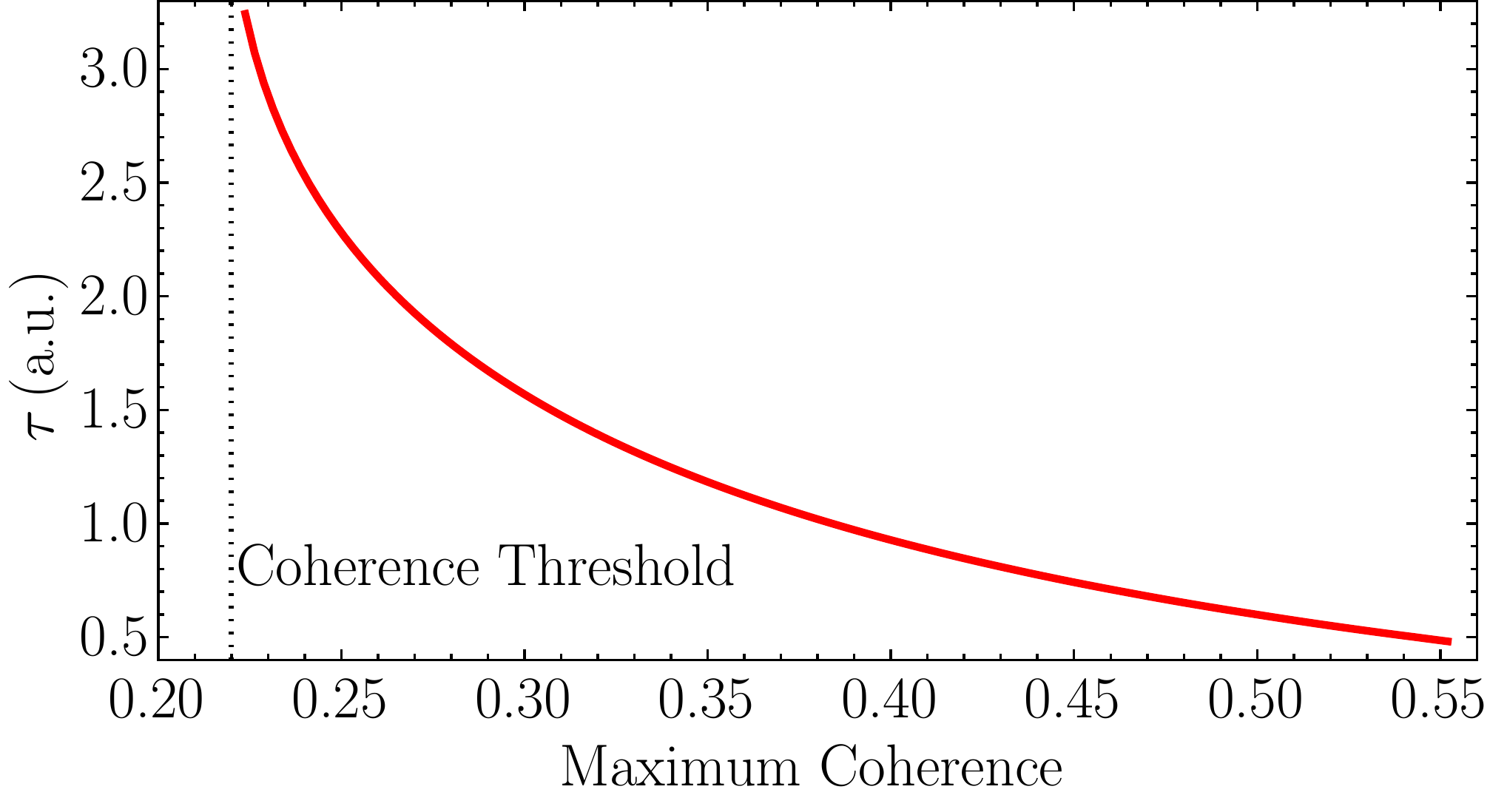}
	
	\caption{(Color online) Intersection time $\tau$ as a function of the maximum coherence. The coupling $\lambda$ varied from $\lambda=0.85$ (a.u.) to $\lambda=3.0$ (a.u.). \label{tauxcoh} }
\end{figure}

\begin{figure}[t!]
	\includegraphics[scale=0.41]{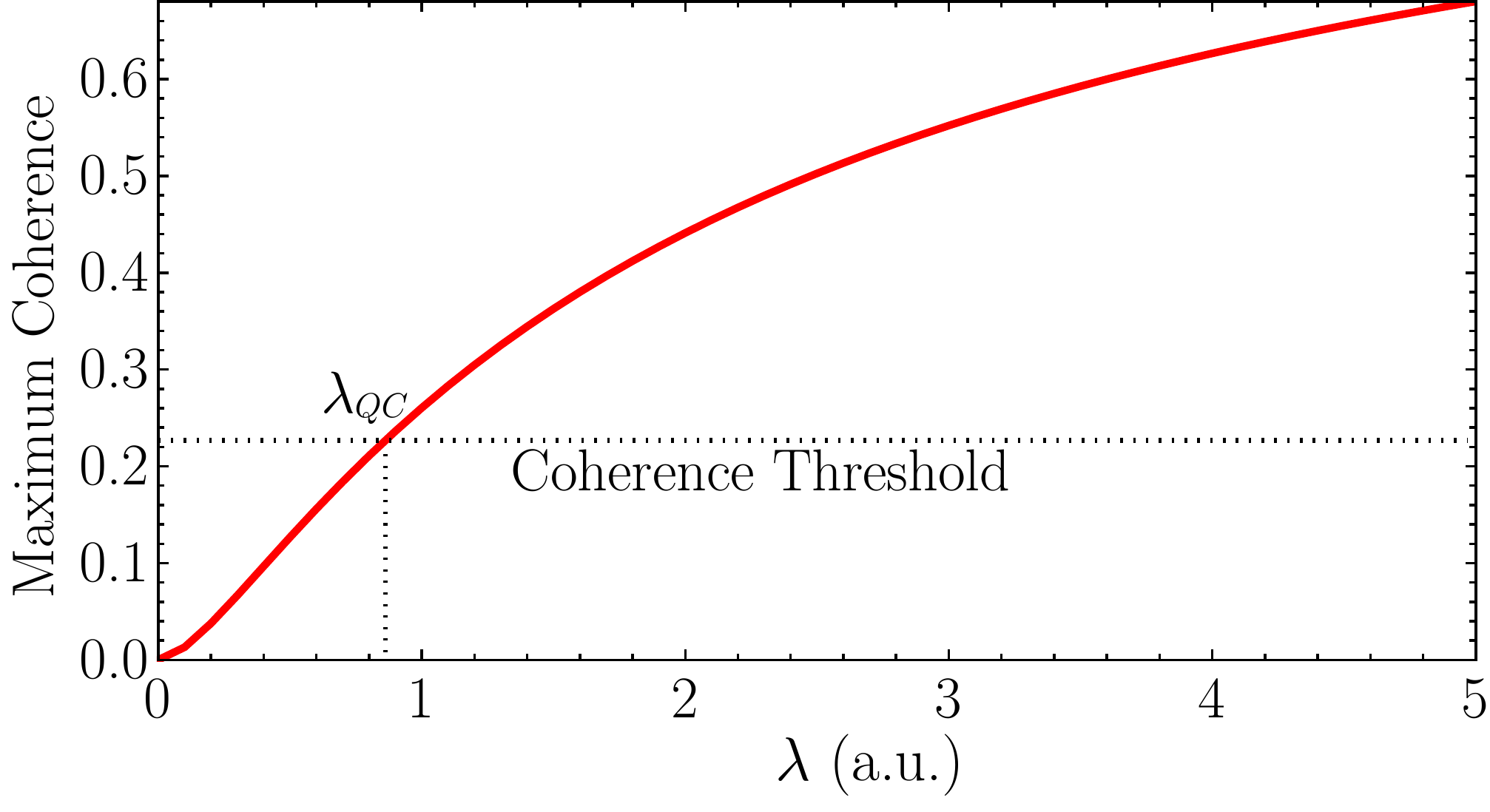}
	
		\caption{(Color online) Maximum coherence attained during the dynamics as a function of the coupling strength $\lambda$. \label{cohxlam}}
\end{figure}

We first consider the system dynamics and the amount of time required to observe any quantum advantage, i.e., for the sink population  $P_Q(t)$  to surpass $P_C(t)$, whenever this is possible. 
In Fig. \ref{dynamics}, we present the time evolution of the sink populations $P_Q(t)$ and $P_C(t)$, as well as the coherence $C(\rho_Q)$, for some values of the intersite coupling strength $\lambda$. 
The top panel on the left shows the interesting case where the quantum and classical scenarios lead to the same efficiency. 
We call $\lambda_{QC}$ the value of $\lambda$ for which this happens. 
The other panels depict cases corresponding to $\lambda>\lambda_{QC}$, for which quantum advantage can be observed. 
It is remarkable that the time required for $P_Q(t)$ to surpass $P_C(t)$ seems to decrease as the global maximum of $C(\rho_Q)(t)$ increases. 
To further explore this feature, let us call this timespan the {\it intersection time} $\tau$. 
Its behavior as a function of the maximum coherence attained in the dynamics for values $\lambda>\lambda_{QC}$ is shown in Fig. \ref{tauxcoh}.
We can see that $\tau$ decreases as this maximum coherence increases.
Moreover, in Fig. \ref{cohxlam}, it is shown that the maximum coherence monotonically increases with $\lambda$.
By considering both Figs. \ref{tauxcoh} and \ref{cohxlam}, it is clear that $\lambda_{QC}$ is unique when the parameters $\gamma$, $\Gamma$, $\Gamma_s$ are fixed.
This means that some quantum advantage will always be observed when $\lambda>\lambda_{QC}$. 
On the other hand, for $\lambda\leq\lambda_{QC}$ we never see any quantum advantage, i.e., $P_Q(t)\leq P_C(t)$ for all times. It is also worth pointing out that these results imply in a  {\it coherence threshold} achieved when $\lambda = \lambda_{QC}$,  above which the quantum scenario is more efficient. For the parameters considered in this simulation, one finds $\lambda_{QC}\approx 0.84$ with a coherence threshold about $0.22$. Interestingly, this is much smaller than the maximum value achieved by  Eq.(\ref{relatentropy}) in the case $N=3$, which is $\ln (8)\approx2.07$ \cite{Baumgratz}.

%

\begin{figure}[t!]
	\includegraphics[scale=0.56]{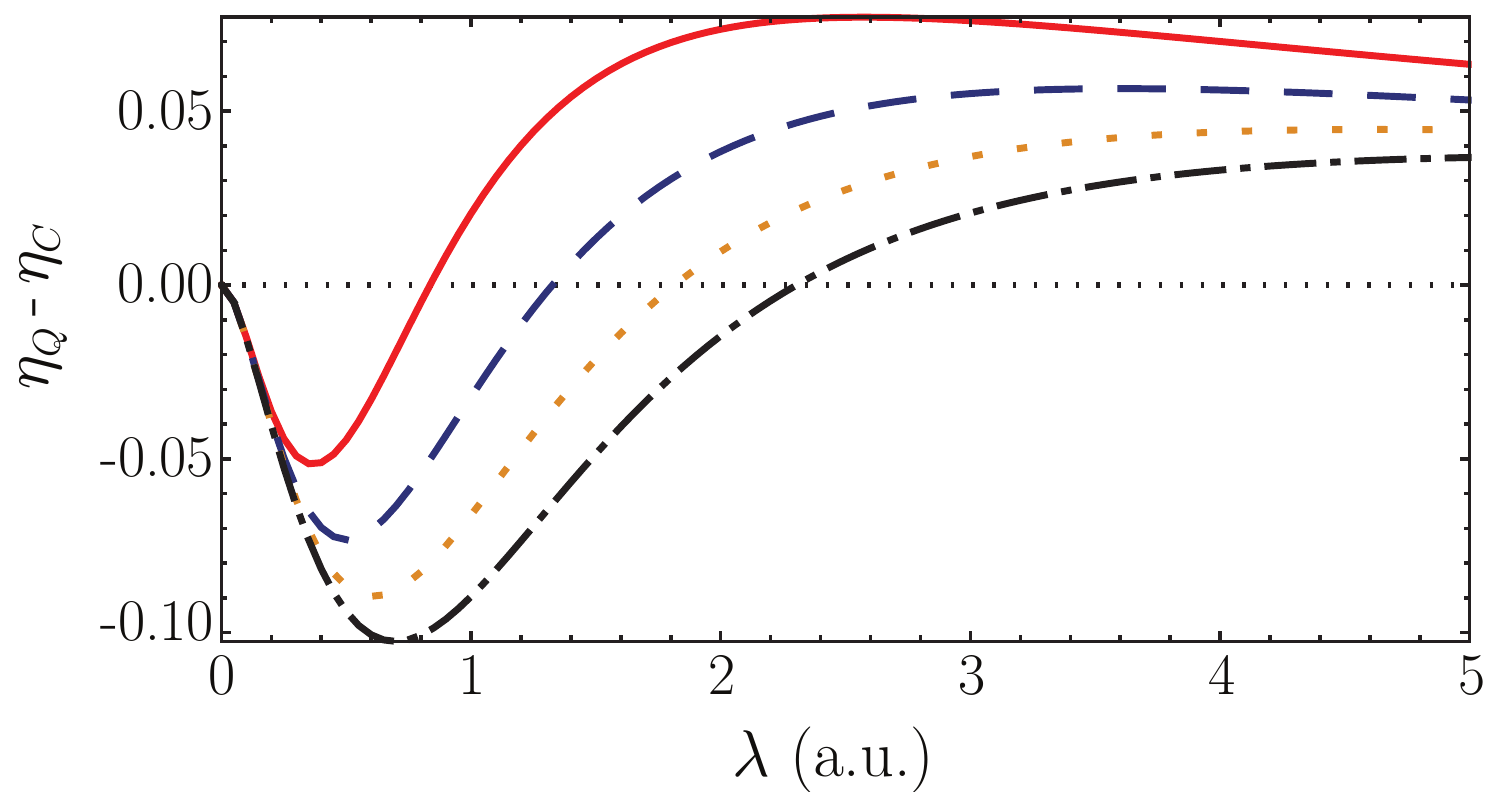}
	\caption{(Color online) Difference between the quantum and classical efficiencies as a function of the site-to-site coupling strength, and for different dephasing rates $\gamma=0.25$ a.u. (red solid line), $\gamma=0.5$ a.u. (blue-dashed line), $\gamma=0.75$ a.u. (orange-dotted line), $\gamma=1.0$ a.u. (Black-dot-dashed line) and $\Gamma=0.5$ a.u.
		\label{result1}}
\end{figure}

\begin{figure}[t!]
	\includegraphics[scale=0.4]{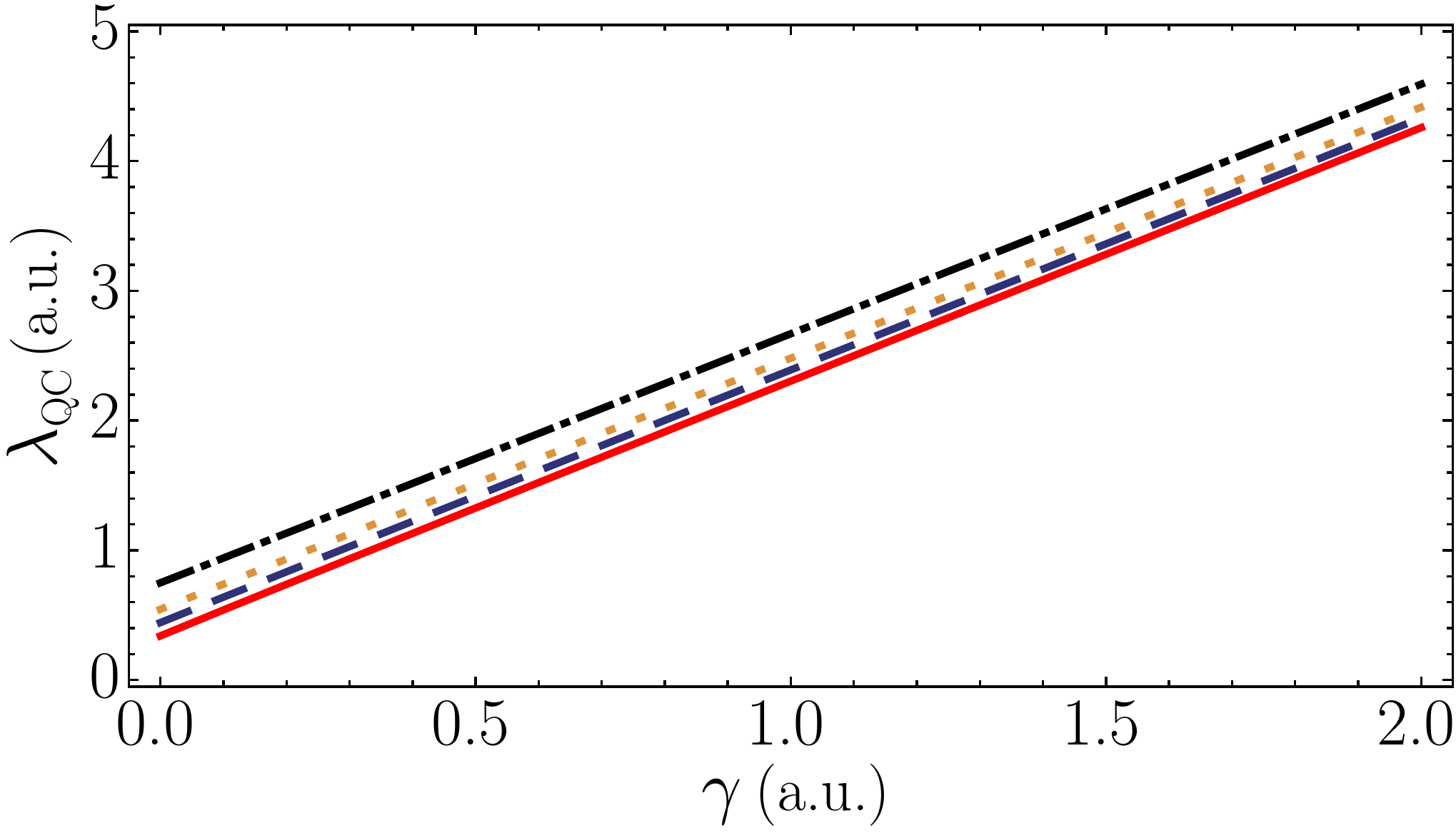}
	
	\caption{(Color online) Coupling $\lambda_{QC}$ as a function of $\gamma$ for different values of dissipation $\Gamma$, $\Gamma=0.5$ a.u. (red solid line), $\Gamma=0.75$ a.u.(blue-dashed line), $\Gamma=1.0$ a.u. (orange-dotted line), $\Gamma=1.5$ a.u. (Black-dot-dashed line)
		\label{mincoupling}}
\end{figure}

 In Fig. \ref{result1}, we plot the difference between the transport efficiencies in the quantum and classical scenarios, $\eta_Q-\eta_C$, as a function of the coupling strength $\lambda$, for different values of dephasing rate $\gamma$ and for a fixed local dissipation rate $\Gamma$. 
 We can see that the quantum scenario is not always more efficient than its classical counterpart, and that there is a particular value for the coupling $\lambda$ for which $\eta_Q = \eta_C$, previously defined as $\lambda_{QC}$ which, in turn, depends on the dephasing rate.
 This dependence is depicted in Fig. \ref{mincoupling}, where $\lambda_{QC}$ is seen to increase with $\gamma$ to compensate for the fact that noise depletes nonclassical features such as quantum coherence.
 In the same plot, it is also possible to see the role played by local dissipation. 
 It is interesting to see that its effect is to shift $\lambda_{QC}$ to higher values for a fixed dephasing rate $\gamma$.
 Once again this is explained by the fact that more noise means less coherence, in general.
 Finally, it is notable that the dependence of $\lambda_{QC}$ with $\gamma$ is basically linear.

%

\subsection{Quantum invasiveness}

The previous discussion addresses the importance of coherence in the dynamics when the energy transport is caused by a Hamiltonian term such as $H_I$ in Eq. (\ref{Qcoupling}).
In particular, based on our the previous analyses, one may wonder  how the total coherence might impact the efficiency of quantum transport.
 This question led us to investigate the role played by the integrated coherence \cite{Shahbeigi, Naikoo},
\begin{align}
I(\Phi_{t_0}^t)=\int_{t_0}^{t}C(\rho)dt,
\label{quantifier}
\end{align}
in the efficiency of the energy migration through the chain to the sink.
In Eq. \eqref{quantifier}, $\Phi_{t_0}^t$ is the time dependent map or operation associated with the quantum master Eq. (\ref{qdynamics}), when the system evolves from an initial time instant $t_0$ to an arbitrary instant $t$. For our purposes, we set $t_0=0$ and $t\rightarrow\infty$.

\begin{figure}[t!]
	\includegraphics[scale=0.58]{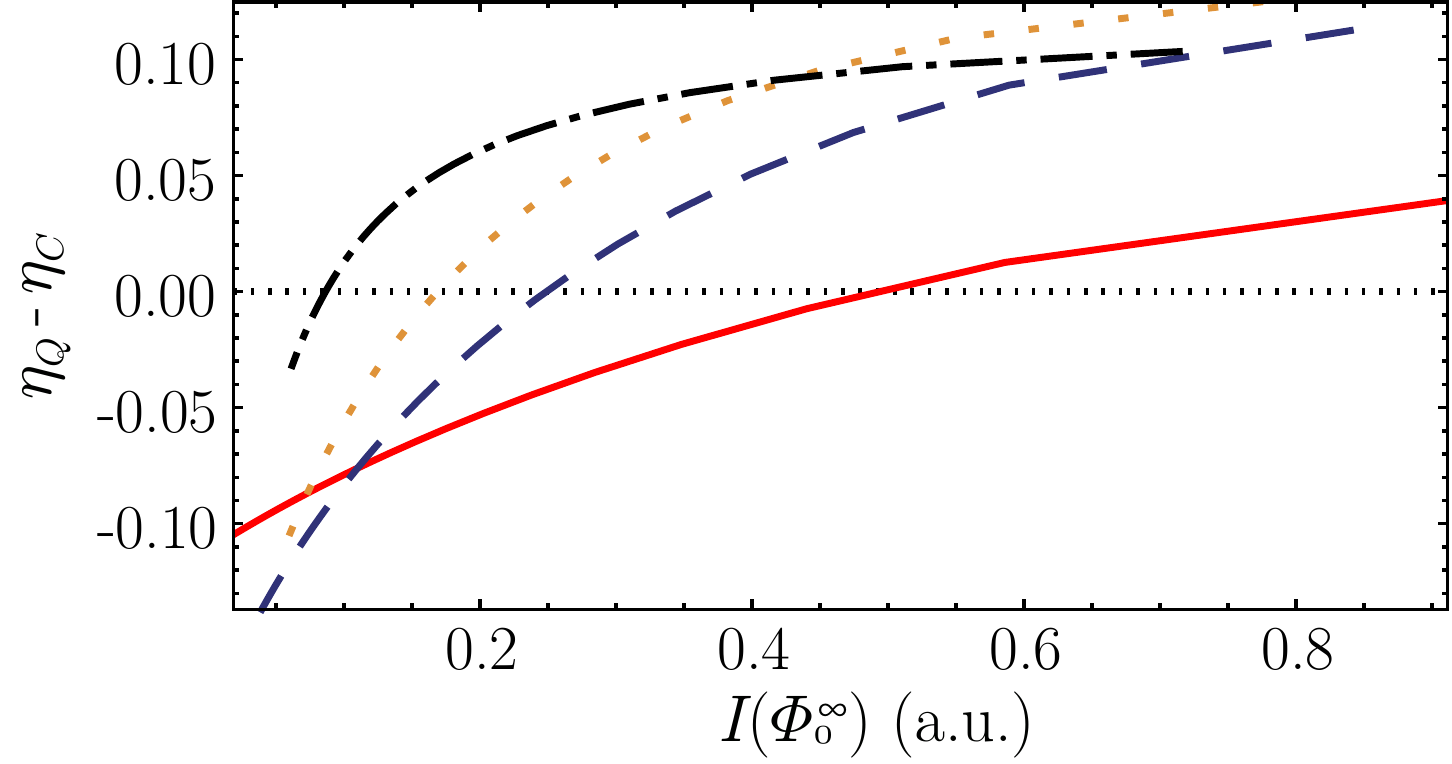}
	\caption{(Color online) Parametric plot showing the difference between the quantum and classical efficiencies as a function of the invasiveness quantifier [Eq. (\ref{quantifier})]. To generate the curves, the dephasing rate parameter $\gamma$ was varied in the interval $[0,2]$. We have used $\lambda=0.5$ a.u. (red solid line), $\lambda=1.0$ a.u. (blue-dashed line), $\lambda=1.5$ a.u. (orange-dotted line), $\lambda=3.0$ a.u. (Black-dot-dashed line) and $\Gamma=0.5$ a.u.
		\label{I}}
\end{figure}

In Fig. \ref{I}, we show a parametric plot of the difference between the quantum and classical efficiencies as a function of the integrated coherence, $I(\Phi_{0}^\infty)$, for different site-to-site couplings $\lambda$. 
To generate those plots, the dephasing rate was varied from  $\gamma=0$ a.u. to $\gamma=2.0$ a.u.
It is remarkable that, for fixed $\lambda$, the quantity $\eta_Q-\eta_C$ increases with $I(\Phi_{0}^\infty)$, indicating that the integrated coherence helps quantum transport. 


In a first moment, we can think of Eq. (\ref{quantifier}) as just the area under the curve of coherence, produced by the map $\Phi_{t_0}^t$.
Nonetheless, we will now show that this quantity is also a nonclassicality quantifier when the quantum operation $\Phi_{t_0}^t$ is seen as a {\it dynamical} resource \cite{Moreira4}.
This resource is essentially different from coherence, which is a property of a given quantum state, not a map.


\begin{figure*}[t!]
	\includegraphics[scale=0.51]{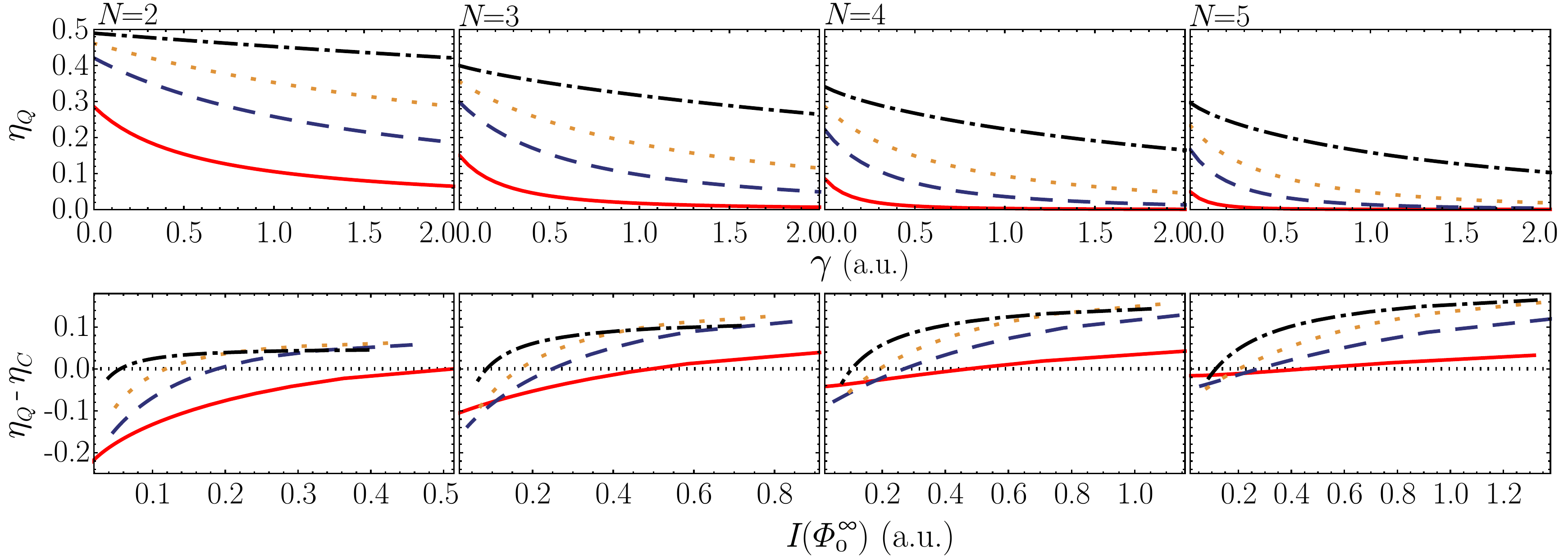}
	\caption{(Color online) In all panels, $\Gamma=0.5$ a.u. is fixed. The top panels show the efficiencies $\eta_Q$ as a function of the dephasing rate $\gamma$ for different values of site-to-site coupling $\lambda$. The bottom panels show $\eta_Q-\eta_C$ as a function of the invasiveness quantifier $I(\Phi_0^\infty)$ for different values of $\lambda$. We have used $\lambda=0.5$ a.u. (red solid line), $\lambda=1.0$ a.u. (blue-dashed line), $\lambda=1.5$ a.u. (orange-dotted line), $\lambda=3.0$ a.u. (Black-dot-dashed line). The number of two-level sites in the chain, $N$, is indicated for each column of panels.}
	\label{difnumber}
\end{figure*}

Let us start with the definition of a quantum invasive operation \cite{Moreira4, Wang, Knee} -- a general quantum operation, represented by a quantum map, is considered to be invasive whenever it disturbs the physical system in a ``nonclassical way.''
To be more precise, everything starts with the definition of classical states  as the eigenstates of a chosen observable $O$, as well as their convex combinations. 
The reasoning here is that, with respect to $O$ measurements, one can assert a classical ontological interpretation to its eigenstates and their convex combinations, as the latter only represent lack of classical information.
Thus, the classical states are called free states.
In turn, free operations must necessarily map a classical state into another classical state, as it is the case with
incoherent completely positive trace preserving maps \cite{Baumgratz}.
A quantifier $I$ of the invasiveness of an operation $\Phi$ with respect to an observable $O$ is expected to satisfy conditions such as positivity, i.e., $ I(\Phi) \ge 0 $ for any physical operation $\Phi$ while $ I(\Phi_\text{Free})=0 $ for any free operation $\Phi_\text{Free}$. 
Additionally, other formal properties such as  monotonicity under free operations
and convexity are demanded - see Refs.  \cite{Coecke,Moreira4} for more details.
 
In our problem, we start by defining the eigenstates of the Hamiltonian $H_F$ in Eq.\eqref{FreeH} and their convex combinations as the classical states, i.e., $O \equiv H_F$. 
In turn, the free operations are those leading classical states into classical states, which, therefore, cannot generate coherence.  
In our case, this is represented by the map $(\Phi_{t_0}^t)_C$ associated with the master equation describing the classical scenario in Eq.(\ref{cdynamics}).
Consequently,  $C[(\Phi_{t_0}^t)_C(\rho_C)] = 0$ for any time interval $[t_0,t]$. 
Therefore, we have that the propagation of a single excitation governed by Eq. (\ref{cdynamics}) will remain classical, which means an incoherent mixture of the eigenstates of $H_F$, and the integrated coherence will be zero.
On the other hand, for a classical initial state -- as it is the case here -- any degree of coherence generated in the quantum scenario will be a result of the quantum invasiveness of $\Phi_{t_0}^t$.
This guarantee that conditions of positivity, monotonicity under the composition with free operations $(\Phi_{t_0}^t)_C$, and convexity are fulfilled.
Such a reasoning leads us to recognize $I(\Phi_{t_0}^t)$ as a quantifier of quantum invasiveness in the transport scenarios described by Eq.(\ref{qdynamics}), where $H_I$ plays an important role.

As a conclusion, quantum invasiveness, here quantified by the integrated coherence generated in the whole dynamics, is a resource that can benefit quantum transport as illustrated in Fig. \ref{I}.  Previously, quantum invasiveness has also find applications in the inference of nonclassicality from nonlinear electronic spectroscopy \cite{2D}, an important experimental technique in the study of energy transfer pathways in complex molecular aggregates \cite{path}.

Finally, we now investigate the effect of the number $N$ of sites on small chains. 
These small chains are within the grasp of current technology as we discuss later on this work.
{For each column of panels in Fig. \ref{difnumber}, we have a chain with $N$ two-level sites, for $N = 2, 3, 4, 5$.
In the top panels, we show the quantum efficiency $\eta_Q$ as a function of $\gamma$ for different values of $\lambda$. 
In general, by increasing  the number of sites, the quantum efficiency decreases.
This can be understood as the effect of the number of local dissipators which, according to our model, also increases with $N$. They drag out the energy of the chain before it reaches the $N$th site. 
For the incoherent and continuous injection of energy, with no local energy dissipators, such dependence does not naturally manifest for the coherent hopping mechanism in  Eq.(\ref{Qcoupling}). It can be restored by using, for example, local dephasing \cite{Manzano3, Manzano, Manzano2}.
Similarly, $\eta_Q$ decreases with $\gamma$, for fixed $\lambda$ in the top panels.
The classical efficiency $\eta_C$ (not shown)  does not depend on $\gamma$, and 
an increase in $N$ also depletes the classical efficiency.




Despite the fact that individually} $\eta_Q$ and $\eta_C$ diminish with $N$, their difference $\eta_Q-\eta_C$ or the quantum advantage becomes more pronounced as $N$ increases. This can be seen from the bottom panels of Fig. \ref{difnumber} when we take a curve with fixed $\lambda$ and observe it being displaced upwards with $N$ (panels from left to right). The main feature contained in the bottom panels of Fig. \ref{difnumber}, however, is the monotonic behavior of $\eta_Q-\eta_C$ with the quantum invasiveness $I(\Phi_0^\infty)$. Once again, for a fixed $\lambda$, $\eta_Q-\eta_C$ increases with the quantum invasiveness for all chain sizes considered. This persistent behavior further suggests the value of quantum invasiveness for quantum transport.



\subsection{Feasibility}

Finally, to illustrate the application of our framework in a physical scenario, we now present a simple example of an experimentally accessible setup involving  coupled two-level systems with engineered site-to-site coupling constants.
In Ref. \cite{experimental}, an experimental scheme to couple two superconducting gatemon qubits is presented. The physical mechanism promoting the indirect coupling between the qubits is basically that of a quantum bus implemented with the help of a detuned bosonic mode coupled to the qubits \cite{old}.
In the experiment, an epitaxial semiconductor-superconductor nanowire is used as a field-effect switch to tune a superconducting cavity. Since the effective coupling between the two qubits is inversely proportional the qubit-cavity detuning \cite{old}, this ability to tune the cavity allowed for induced  coupling strengths ranging from $0$ to about $25$ MHz. The direct capacitive coupling between the qubits was estimated to be less than $1$ MHz. The qubit frequencies of the resonant qubits were around $5$ GHz, which truly justifies the local bath assumption used in our approach.

\begin{figure}[t!]
	\includegraphics[scale=0.57]{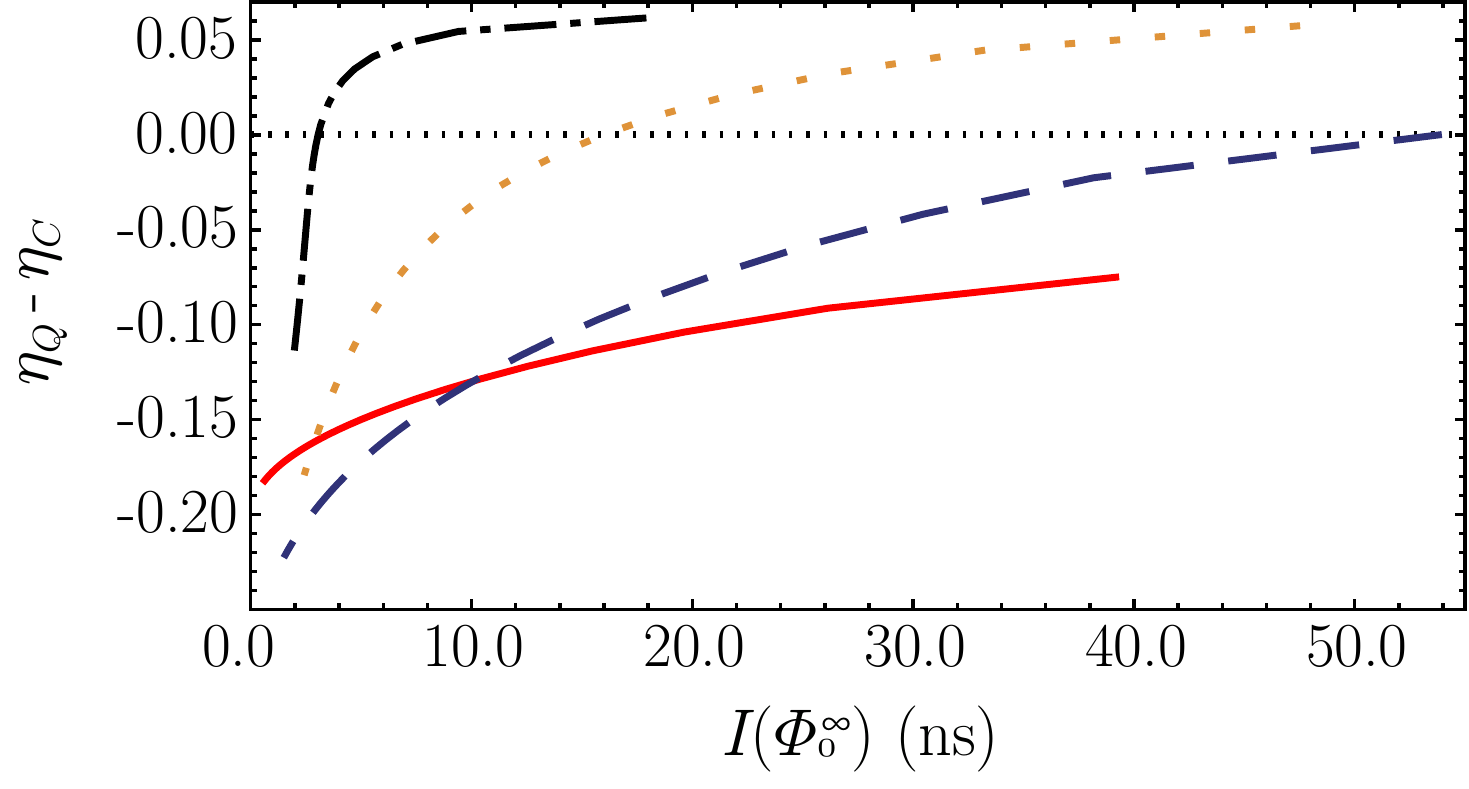}	
	\caption{(Color online) Parametric plot with accessible experimental parameters  \cite{experimental} showing the difference between the quantum and classical efficiencies as a function of the invasiveness quantifier [Eq. (\ref{quantifier})]. Here, the units of $I(\Phi_0^\infty)$ are in nanoseconds (ns).  To generate the curves, the dephasing rate  $\gamma$ was varied from $\gamma=0$ MHz to $\gamma=20$ MHz. We kept $\Gamma=5.0$ MHz and $\Gamma_s=10.0$ MHz. We have used $\lambda=2.5$ MHz (red solid line), $\lambda=5.0$ MHz. (blue-dashed line), $\lambda=10.0$ MHz, (orange-dotted line), $\lambda=15.0$ MHz (black-dot-dashed line).}
	\label{exp}
\end{figure}

We used the range of parameters reported in \cite{experimental} to produce the curves shown in Fig. \ref{exp} for the case $N=2$.  Interestingly, the transition from classical to quantum in terms of the values assumed by $\eta_Q-\eta_C$, as well as its monotonic behavior with quantum invasiveness, are all within the available range of coupling strengths as reported in \cite{experimental}.

\section{Conclusion}\label{Conclusions}

To summarize, we focused on the relationship between nonclassical resources, such as coherence and quantum invasiveness, and transport efficiency in coupled quantum systems.
By defining the set of classical states as the eigenstates of the decoupled system Hamiltonian and their convex combinations, as well as classical transport operations, we constructed a classical transport scenario.
This is used as a benchmark in the comparison with the quantum scenario, where coherent couplings between the sites of the chain are allowed.
To illustrate our framework, we focused on a linear chain of two-level systems.
By using the relative entropy of coherence as a coherence quantifier, we investigated how this resource is associated with an advantage in terms of transport efficiency, when compared to the classical scenario. 
We then focused on the role played by the integrated coherence, which we showed to be a quantifier of the invasiveness of the quantum operation associated with the quantum dynamics. 
Then, we were able to investigate how quantum invasiveness positively impacts the efficiency of quantum transport.

It is important to remark that there are other interesting approaches to characterize nonclassicality in quantum transport. 
For example, in Ref. \cite{FNori}  nonclassicality is characterized by the violation of the Legget-Garg inequality and in Refs. \cite{quantuness2,quantuness3} by the distance with respect to a set of classical states.  It is also worthwhile to remark that  symmetric linear chains such as the ones studied here are not prone to dephasing assisted transport \cite{PlenioTransport, Guzik, Elinor}.
Given the importance of transport for quantum technologies and biological molecular systems presenting quantum coherence   \cite{PlenioTransport,Guzik,poly,wavelike,Elinor,Schachenmayer,Semiao,Feist,EngVibAssEnTranfIons,fmo1,fmo2,fmo3,EnvoiAssTranspIonsMaier,NMTransport,Cao,lhs1,lhs2,lhs3,solarcell,EET}, we believe that our approach, which characterizes nonclassicality based on a resource theory for quantum operations \cite{Moreira4}, will help to shed some light on the implications of quantumness for transport.

\section*{ACKNOWLEDGMENTS} 
I. M. and S. V. M. acknowledge financial  support from the Brazilian agency Coordena\c{c}\~ao de Aperfei\c{c}oamento de Pessoal de N\'ivel Superior (CAPES).
F. L. S. acknowledges partial support from the Brazilian National Institute of Science and Technology of Quantum Information (CNPq-INCT-IQ 465469/2014-0), CNPq (Grant No. 305723/2020-0) and CAPES/PrInt – Process No. 88881.310346/2018-01.

\end{document}